\documentclass{edp-conf}
\usepackage{graphicx}
\begin{document}

\TitreGlobal{SF2A 2003}

\title{Supermassive black holes (SMBH) and formation of galaxies}
\author{Combes, F.}
\address{LERMA, Observatoire de Paris, 61 Av. de l'Observatoire, F-75014, Paris, France}
\runningtitle{Supermassive black holes}
\setcounter{page}{237}
% Keep this line, even if the page will be settled afterwards..
\index{Combes, F.}

\maketitle
\begin{abstract} 
The recently confirmed correlation between the mass of SMBH and bulges of galaxies (and
their central velocity dispersion), suggest a common formation scenario for galaxies 
and their central black holes. 
Common fueling can be invoked through internal dynamical processes, external accretion, and
hierarchical merging of structures. The success of recent theories is reviewed, as the
self-regulated growth of both bulges and SMBHs, the predicted AGN statistics, when activity is
triggered by accretion and mergers, the predicted frequency of binary SMBH and consequences. In
particular, the SMBH growth problem can now be revised, invoking intermediate-mass black holes
(IMBH) as BH seeds in the early universe. As a by-product, the merger of
binary SMBHs help to heat and destroy central stellar cusps. Remaining problems are mentioned. 
\end{abstract}
%
%%-----------------------------
%%      your text
%%-----------------------------
\section{Introduction}
  A major advance in this topic in recent years has been
the determination of the Black hole to Bulge mass Relation
(cf Fig. 1; this will be called in the following BBR; 
Magorrian et al 1998, Gebhardt et al 2000, Merritt \& Ferrarese 2001, Shields et al 2003).
The determination of the BBR has been made by various methods:
1. stellar proper motions for the Galactic center BH (Sch\"odel et al. 2003, Ghez et al. 2003),
2. stellar absorption lines, to obtain the stellar kinematics,
3. ionised gas emission lines (less reliable, since affected by outflows, inflows), and
also masing gas emission lines,
4. reverberation mapping, exploiting time delays between variations of AGN continuum,
and broad line emission, giving the size of the emitting gas region, combined with
the gas Doppler velocity to give the virial mass (Peterson \& Wandel, 2000)
5. ionization models: method based on the correlation between
quasar luminosity and the size of the Broad Line Region (BLR, Rokaki et al 1992).

Some progress has also been made in the search of intermediate mass black holes
(IMBH), for example in the globular clusters M15 in our Galaxy and G1 in M31:
in M15, the mass of the central object is lower than 10$^3$ M$_\odot$ and could be
stellar remnants (van der Marel 2003), while in G1, a BH of 2 10$^4$ M$_\odot$
is identified, and obeys the BBR (Gebhardt et al 2002).

\begin{figure}[h]
   \centering
   \includegraphics[width=9cm]{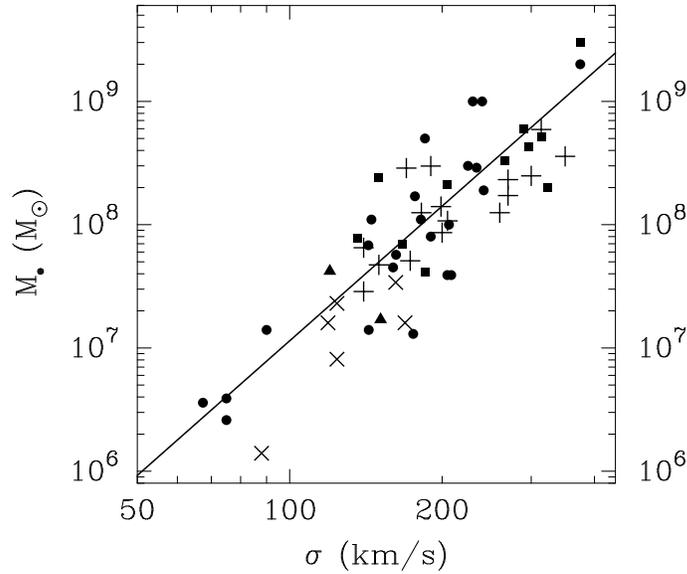}
      \caption{The BBR: BH mass versus the velocity dispersion $\sigma$
inside the effective radius of the bulge. Filled circles indicate BH mass measurement
from stellar dynamics, squares from ionized gas, triangles on maser lines,
crosses are from reverberation mapping, and ``plus signs'' from ionization models
(from Kormendy \& Gebhardt 2001).}
       \label{fig1}
   \end{figure}

The demography of SMBH is now much better known. It was already
suspected that AGN were active during a
short duty cycle of $\sim$ 4 10$^7$ yr,
and that many galaxies today should host a starving
black hole (Haehnelt \& Rees 1993). The observed BBR now
strongly constrains the duty cycle time-scale. Also the cosmic background radiation
detected at many wavelengths constrains the formation
history. 
The volumic density of massive black holes today is derived,
from the observed density of bulges, and the 
proportionality factor M$_{BH}$ = 0.002 M$_{bulge}$  (BBR).
And independently, the light that should have been radiated at
the formation of these BHs can be computed, redshifted and compared to the observed
cosmic background radiation: in the optical, we see only 10\% of the expected
flux, but 30\% in X-rays, and 80\% in the infra-red. 
The accretion radiation does not get out in optical light,
probably due to the extinction.

\section{BH growth}

Since AGN are observed early in the universe, with
powerful emission, indicating very high BH masses
(higher at z=3 than today), 
their growth time-scale is a problem.

\smallskip\noindent
{\bf 2.1  Quantifying the problem}\\
\smallskip\noindent
To have an order of magnitude, 
and simple dimensional relations,
let us assume spherical accretion, 
from an accretion radius R$_{acc}$ = 0.3 M$_6$/v$_2^2$ pc,
where M$_6$ is the mass of the BH in 10$^6$  M$_\odot$, and
v$_2$ the velocity in 100km/s (corresponding to the effective stellar velocity
inside the galaxy nucleus, related to the bulge mass). The
canonical Bondi accretion rate is then:
dM/dt = 4 $\pi$ R$^2$ v $\rho$ = 10$^{-4}$ M$_\odot$/yr M$_6^2$/v$_2^3$ $\rho$,
where $\rho$ is the local density in M$_\odot$/pc$^3$.

Since dM/dt $\propto$ M$^2$, then the accretion time is $\propto$ 1/M,
t$_{acc}$ $\sim$ 10$^{10}$ yr/M$_6$ v$_2^3$/$\rho$;
for very low mass BH, this takes much larger than the Hubble time.
Therefore the formation of SMBH requires a large seed, mergers of BH, or very large
densities, like in the Milky Way nucleus, 10$^7$ M$_\odot$/pc$^3$.

If these conditions are fulfilled, the growth of massive BH can then 
be accretion-dominated, i.e. t$_{growth}$ = t$_{acc}$. This phase could
correspond to moderate AGN, like Seyferts, and the luminosity is
increasing as  L $\propto$ dM/dt $\propto$  M$^2$. At some point,
the luminosity will reach the Eddington luminosity, since L$_{edd} \propto$ M.
The Eddington ratio increases as 
L/L$_{edd} \propto$ M, the BH growth slows down when approching L$_{edd}$,
corresponding to a QSO phase. The time-scale of this powerful
AGN phase is
t$_{edd}$ = M/(dM/dt)$_{edd}$ = 4.5 10$^7$ yr (0.1/$\epsilon$)
(where $\epsilon$ is the usual radiation efficiency).
Equating t$_{acc}$ = t$_{edd}$, this occurs for 
M = 2 10$^8$ M$_\odot$ v$_2^3$/$\rho$ ($\epsilon$/0.1).
Wang et al. (2000) propose that tidal perturbations help to grow
a SMBH from a small seed, by boosting the accretion, and then
lead to the BBR.
 
\smallskip\noindent
{\bf 2.2 Are the BH the first objects to form?}\\
\smallskip\noindent
One solution to the BH growth problem would be that massive BH
form very early at high redshift, as the remnants of Pop III stars.
Fragmentation must be inefficient, and the first stars
could reach $\sim$ 300M$_\odot$, since radiative losses are
negligible at zero metallicity. 
Above 260 M$_\odot$, the objects could collapse to a BH directly
(Madau \& Rees 2001, Schneider et al. 2002). In almost all mini-haloes,
10$^5$ M$_\odot$ IMBH could be formed, by the merging of these seeds.

\smallskip\noindent
{\bf 2.3 Do IMBH exist?}\\
\smallskip\noindent
Some evidence for the existence of 
IMBH would be welcome, to support theories. However, their observation
is very difficult, both by the kinematics, since their gravitational influence
is small, and from their possible AGN activity, since the expected
luminosity is weak. According to the extrapolation of BBR, these IMBH
should be searched as AGN in dwarf galaxies: a good
candidate is NGC 4395 (Filippenko \& Ho 2003), where the BH mass
is likely to be 10$^4$-10$^5$ M$_\odot$ (radiating much below
the Eddington limit).
The problem of this search is that dwarf galaxies
frequently host nuclear star clusters of $\sim$ 10$^6$ M$_\odot$,
hiding the weak AGN. They are best observed in the
Local Group; a famous example, M33, does not host any BH
more massive than  10$^3$ M$_\odot$, which is already 10
times below the value expected from the BBR.

\smallskip\noindent
{\bf 2.4 Double BH in the Milky-Way nucleus}\\
\smallskip\noindent
Evidence for an IMBH could come from the Milky Way nucleus:
Hansen \& Milosavljevic (2003) propose its existence to explain
the observation of 
bright stars orbiting within 0.1pc, which are are young
massive main-sequence stars, in spite of an environment hostile
to star-formation. Aternative solutions could be
star mergers, or exotic objects (Ghez et al. 2003).
In the IMBH scenario, stars were formed in a star cluster 
outside the central pc, and then dragged in by
a BH of 10$^3$-10$^4$ M$_\odot$. The decay time-scale by
dynamical friction for normal
stars is too large (much longer than the massive stars life-time),
but for the IMBH, this time-scale is 1-10 Myr.
Stars may be dragged inwards even after the star cluster 
has been disrupted.

Such a system SMBH-IMBH and a gas disk may reveal interesting
dynamics; it is  similar to a protosolar system,
with the Sun-Jupiter couple, resonant effects like
planetary migration are expected (Gould \& Rix 2000).

\section{Interpretation of the BBR}

Several models have been proposed to account
for the BBR, all involving a simultaneous formation
of bulges and SMBH, and constraining the
feedback processes.

\smallskip\noindent
{\bf 3.1 Feedback due to QSO outflows}\\
\smallskip\noindent
QSO and stars main cosmic formation epoch coincide
(e.g. Shaver et al. 1996). Their common formation
could be regulated by each other, and the QSO outflows
prevent star formation (Silk \& Rees 1998).
The condition for the wind to be powerful enough to
give escape velocity to the gas constrains
the BH mass to M$_{BH}$ $\propto$ $\sigma^5$,
which from the Faber-Jackson relation, gives  M$_{BH}$
 $\propto$  M$_{bulge}$.
But the phenomenon is assumed spherical, 
in reality jets are collimated, the gas is clumpy, 
and compressed to form stars.

\smallskip\noindent
{\bf 3.2 Sinking of Super Star Clusters (SSC)}\\
\smallskip\noindent
Sinking of SSC in a dark halo has been proposed to
form bulges (Fu et al. 2003); the merging of small BH associated
to clusters would provide a mass ratio
M$_{BH}$/M$_{bulge}$ = 10$^{-4}$ only, slightly
below what is observed.

\smallskip\noindent
{\bf 3.3 Radiation drag}\\
\smallskip\noindent
Bulge stars can drive accretion by radiation drag
on the ISM, in  extracting angular momentum
(Umemura 2001).
The M$_{BH}$/M$_{bulge}$ is then an universal constant
depending only on the energy conversion efficiency for
 nuclear fusion of hydrogen to helium.
The efficiency falls as 1/$\tau^2$, with $\tau$ the optical depth of the gas. 
But star formation occurs in a clumpy medium.
Today this mechanism is inefficient, since elliptical galaxies
and bulges have no gas.

\smallskip\noindent
{\bf 3.4 Hierarchical models of galaxy formation}\\
\smallskip\noindent
Hierarchical models explain very well the BBR
(Haehnelt \& Kauffmann 2000).
The scatter is due to:
1.-- M$_{gas}$ of the bulge progenitor depends on $\sigma$, but not on
the formation epoch of the bulge, while M$_*$ depends on both; 
2.-- mergers move the galaxies on the M$_{BH}$- $\sigma$  relation,
even at the end, when there is only BH mergers 
(and not enough gas left to grow the BH).

The gas fraction in galaxies falls from 75\% at z=3 to 10\%  at z=0.
The gas fraction in major mergers is higher in 
fainter spheroids that form at high z, which are more concentrated.
Elliptical/spheroids forming recently have smaller BH.

Typically a seed BH of 10$^6$ M$_\odot$ forms at 5 $<$ z $<$ 10 
and then gas is accreted. For a typical SMBH,
about 30 BH are merged. Today big elliptical's BH accrete
only by merging with small BH, but in the past, gas accretion
was dominant.

\smallskip\noindent
{\bf 3.5 Bar torques}\\
\smallskip\noindent
Bars concentrate mass towards the center, and 
are able to form bulges and fuel a central BH
(e.g. review in Combes 2001).
Bars are also self-regulated:
as soon as 5\% of the galaxy mass is concentrated in the  
nuclear region, the bar is destroyed (in 2-5 Gyr). The
amplitude of the torque is confirmed by observations.
However, to explain the high frequency of bars today,
galaxies have to accrete external gas, leading
to bar renewal (Bournaud \& Combes 2002, Block et al. 2002).

A galaxy is in continuous evolution, 
and accretes mass all along its life.
Several bar episodes can process in a Hubble time
(cf Fig. 2). At each bar episode, both bulge and BH
grow in a similar manner, which explains the BBR.

\begin{figure}[h]
   \centering
   \includegraphics[width=9cm]{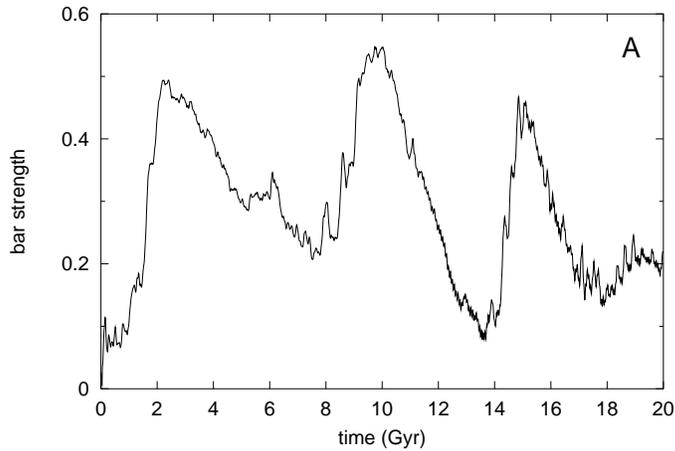}
      \caption{Evolution of bar strength in a simulated
spiral galaxy with gas accretion: several bar episodes
provide a flow of matter towards the center, to fuel
both the bulge and a central blaxk hole (from Bournaud \& Combes 2002).}
       \label{fig2}
   \end{figure}

\section{Cusps and Cores in elliptical galaxies}

The presence of SMBH in every spheroids, and the existence of the BBR,
may explain the observe dichotomy between massive and small ellipticals:
1.-- Cusps (steep power-law in stellar central density profile) are characteristic
of low-mass ellipticals, with disky isophotes and weak rotation;
2.--cores (flat central density profile) are found in high-mass galaxies, 
with boxy isophotes and no rotation. 
Adiabatic growth of a black hole, from gas accretion, and destruction of nearby
stars, produces a cusp (Cipollina \& Bertin 1994).
However galaxy mergers, leading to binary black-holes, and
heating of the central stellar system, are able to produce
the observed cores  (Ebisuzaki et al 1991, Milosavljevic \& Merritt (2001).

\section{Remaining problems}

The expected blue luminosity of AGN, corresponding to the BH growth
and to the BBR, is too large compared
to observations, and models have tried to lower the
radiating efficiency (ADAF, CDAF, ADIOS, extinction).
The expected number of binary BHs from the hierarchical scenario
of galaxy formation is not observed, and mechanisms to
merge them more efficiently have to be found. More questions
remain, as why are the disks so irrelevant in the BBR,
or whether the BBR is already established at high z.
The existence of IMBH has to be proven, 
and the threshold for the BH seeds to be precised. 
Exceptions to the BBR, like M33, have to be searched 
and understood.

%%-----------------------------
%%      your bibliography
%%-----------------------------

\end{document}